\documentclass{article}
\usepackage[dvipsnames]{xcolor}
\usepackage[colorlinks=true, allcolors=Bittersweet]{hyperref}
\usepackage[utf8]{inputenc}
\usepackage[T1]{fontenc}
\usepackage{amsmath}
\usepackage{amssymb}
\usepackage[utf8]{inputenc}
\usepackage{amsthm}
\usepackage{dsfont}
\usepackage[style=apa, natbib=true, backend = biber]{biblatex}
\usepackage{graphicx}
\usepackage{pgfplots}
\usepackage{pgf}
\usepackage{caption, booktabs}
\usepackage{caption}
\usepackage{comment}
\usepackage{float}
\usepackage{subcaption}
\usetikzlibrary{positioning, arrows.meta}
\usepgfplotslibrary{fillbetween}
\usepackage[toc,page,title]{appendix}
\usepackage[a4paper, total={6in, 8in}]{geometry}
\usepackage{setspace}
\begin{document}
\begin{titlepage}
	\center
	\Large{Conflicts and the New Scramble for African Resources: A Shift-Share Approach}
		
	 \vspace{0.5cm}		

\renewcommand*{\thefootnote}{\fnsymbol{footnote}}
  
 \large Raphaël Boulat\footnote{MSc EME - Department of Economics, London School of Economics and Political Science (LSE)}  \\
 	 \vspace{0.2cm}		

	{\large \today}\\[1cm]

  \textit{Extended Essay - Quantitative Economics (EC475)}\footnote{{This Extended Essay is submitted as a requirement for the Quantitative economics module EC475. I want to thank my supervisor, Professor Michael Gmeiner, and Professor Xavier Jaravel, for the extensive advices and help provided over the last few months. I also thank Patrick Leitloff and Prasshanth Sagar, as well as my parents, girlfriend, and friends for all the instructive discussions and emotional support throughout this process. }} \\

\vspace{0.5cm}		

\begin{abstract}
\noindent This paper estimates the causal effect of mineral trade on conflicts in Africa using a Shift-Share IV approach based on an exogenous price-commodity shock. The main result is that an increase in mineral trade significantly increases the number of conflicts while it has no clear effect on fatalities. Exploring heterogeneous effects, I find that a specific group of minerals, oil and fuels, drives the results on the number of conflicts. Moreover, the group of rare minerals such as coltan, precious metals or cobalt has no effect on the number of conflicts but appears to have an important impact on the number of fatalities.
\end{abstract}

\vspace{0.5cm}		

\end{titlepage}

\newpage
 \renewcommand*{\thefootnote}{\arabic{footnote}}
 \setcounter{footnote}{0}
\tableofcontents
\thispagestyle{empty}

 \renewcommand*{\thefootnote}{\arabic{footnote}}
 \setcounter{footnote}{0}
\thispagestyle{empty}
\newpage
\setcounter{page}{1}
\pagenumbering{arabic}

\section{Introduction}

A new type of African natural resources are becoming increasingly valuable in a technology driven world where a critical subset of minerals essential for energy and battery technologies experience rapid growth in demand notably due to the shift toward CO2-neutral policies \citep{jowitt2021battery}. Several external forces therefore compete for such global commodities, creating a new wave of the Scramble for Africa.  \\

\noindent Historically, the 19th century scramble for african resources effectively started with the Berlin Conference (1884-1885) and the partitioning of Africa among European countries. Fragmentation of Africa had long term effects notably on political boundaries \citep{griffiths1986scramble} and on ethnic partitioning, implying higher incidence, severity and duration of political violence \citep{michalopoulos2016long}. Interestingly, this new wave of scramble for Africa involves non-european actors such as America or Asian countries, in particular China \citep{okeke2008second}. This worldwide demand and rising interest for Africa's natural resources raises the stakes and hence potentially conflicts in the region. Literature has shown that mining activity increases the incidence of conflicts at the local level subsequently disseminating violence across broader territories \citep{berman2017mine, de2020origin}. \\  

\noindent To measure the social cost of this new scramble for African minerals, this paper investigates the impact of mineral trade on conflict in Africa between 2000 and 2020 using a SSIV methodology based on an exogenous price-commodity shock weighted by mineral industry shares reflecting heterogeneous shock exposure \citep{borusyak2022quasi}. From an economic theory perspective, the hypothesis investigated is that higher demand for minerals from the global market because of several factors (climate change related or not) increases conflict through higher incentives to fight and scramble for scarce resources. An additional question of interest is to know whether existing conflicts become more fatal if the stakes are higher.  \\

\noindent The contribution of this study is twofold. First, applying a SSIV methodology for the first time in this context enables to evaluate whether the pretty well-established result that a rise in natural resources trade increases conflict is robust to this method. Most importantly, the structure of the data allows to investigate which type of mineral trade matters for two conflict indicators (number of conflicts and fatalities), which provides policy relevant information, notably in the context of climate change. Indeed, this study aims at providing a comprehensive and systematic presentation of which type of mineral matters for conflict in Africa, a question that has not been extensively covered in the literature. Understanding which minerals play critical roles in the stability of resource-rich nations could serve as an additional factor to consider when designing future environmental policies. \\

\noindent Matching trade data at the country level (UN Comtrade) and conflict data from the Armed Conflict Location Events Data (ACLED), I find that higher mineral trade has a causal effect on the number of conflicts but not on number of fatalities. This result is robust to different specifications and falsification tests. When it comes to which type of minerals matters, it seems that only mineral fuels/oils such as gas or
petrol matter for the number of conflicts. Moreover, rare minerals such as coltan, precious metals or cobalt have no effect on the number of conflicts but have an important impact on the number of fatalities, potentially  consistent with the hypothesis that
growing demand for this type of rare resources leads to an escalation in violence and the degree of mortality of conflicts.   \\

\noindent The remainder of this paper is organized as follows. Section 2 presents the political background and Section 3 discusses the data used. Section 4 introduces the methodology and Section 5 the main results. Finally, I present the heterogeneity analysis in Section 6, highlight some limitations in section 7  while  Section 8 concludes.

\begin{section}{Background}

Political theory has long been intrigued by the interplay between trade and conflict. In 1748, Montesquieu \citep{de1844esprit} offered an influential perspective on this question, positing that trade heightens the opportunity cost of war and therefore provides an incentive to minimize military conflicts. Conversely, a contrasting  ``realist view'' argues that increased economics rivalry between nations tends to intensify political tension \citep{copeland1996economic, thoenig2023trade}. \\

\noindent Economic theory soon overtook the topic and an extensive literature exists on the question. In 1980,  Montesquieu's theory has been explored in an economic model finding that mutual dependence established between two trading partners is sufficient to raise the costs of conflict, diminishing levels of dispute \citep{polachek1980conflict}. Later, \cite{martin2008civil} found that a pair of countries with more bilateral trade has lower probability of bilateral war. However, the authors argue that multilateral trade openness has the  opposite impact: countries that engage more importantly in global trade face a heightened likelihood of conflict. This arises because increased multilateral trade openness reduces their relative dependence with any specific bilateral partner, thereby lowering the opportunity cost of engaging in military hostilities. A large body of the literature also examines the impact of conflict on trade \citep{keshk2004trade, martin2008civil, martin2008make, glick2010collateral}, generally finding that conflicts are disruptive for trade. \cite{glick2010collateral} find that trade is substantially reduced not only between belligerent nations, where trade drops by 85\%, but also with nations that are not directly involved in the conflict in which trades decreases by 12\%. \\

\noindent On top of the political and classical economic theory perspectives on trade and war, the link between conflict and specific resources such as natural riches have been of particular interest in more recent empirical economics and policy debates. Indeed, conflicts play a prominent role amongst the several channels proposed to explain the concept of resource curse, which is a paradox that countries rich in non-renewable natural resources tend to display poor economic performance \citep{maystadt2014mineral, bruckner2012oil, ploeg2011natural}. Violence has been shown to respond to economic incentives created by mineral resource extraction \citep{berman2017mine, de2020origin}. Natural resources have also been found to extensively matter for organized crime \citep{buonanno2015poor}, homicide \citep{couttenier2017wild}, interstate war \citep{caselli2015geography} or mass killing of civilians \citep{esteban2015strategic}. The literature has also investigated specific commodities such as oil \citep{ross2006closer, lei2014giant, cotet2013oil}, which paves the way to more commodity-specific comparisons. This paper goes further in this direction analyzing different categories of minerals.     \\ 

\noindent The latest conflict happening at the border between East DRC and Rwanda is an example highlighting that this question is still highly relevant. Nicolas Kazadi, DRC’s finance minister, said in March 2023 that Rwanda exported close to one billion dollars in resources\footnote{Resources include gold, tin, tantalum, and tungsten} last year, even though the country has few mineral deposits of its own\footnote{\url{https://www.ft.com/content/ecf89818-949b-4de7-9e8a-89f119c23a69}}. In a region characterized by persistent instability where more than 6 million people died since 1996 due to repeated conflicts, and nearly 7 million individuals have been displaced internally, as reported by the United Nations\footnote{\url{https://www.france24.com/en/africa/20231030-record-6-9-million-internally-displaced-in-dr-congo-un-says}}, the abundance of natural resources—especially precious minerals—found in Congolese soil has spread the conflict in eastern DRC\footnote{\url{https://www.cfr.org/global-conflict-tracker/conflict/violence-democratic-republic-congo}}. \\

\noindent One important motivation of this paper lies behind the fact that use of mineral resources reached an unprecedented level and demand will probably continue growing in the coming decades \citep{vidal2021modelling}. Challenges regarding the adequacy of mineral resources in the context of population expansion and increasing standards of living have been a concern since the end of the 18th century, as highlighted in the work of  \cite{RePEc:hay:hetboo:malthus1798}. Even though research has shown that predictions of peak production due to resource exhaustion in the next 20–30 years are not necessarily correct \citep{meinert2016mineral}, the aim for a carbon neutral economy by 2050\footnote{\url{https://climate.ec.europa.eu/eu-action/climate-strategies-targets/2050-long-term-strategy_en}} comes with new challenges. Building new facilities such as wind turbines or solar power stations requires vast amounts of metals and other raw materials \citep{vidal2013metals}. This objective also comes with the promotion of new energy vehicles and it has been found that the rapid development of electric vehicles (EVs) contributed to a dramatic increase in in-use metal stocks from 0.7 kt in 2009 to 1.1 Mt in 2019 \citep{yang2021uncovering}. The higher scarcity and demand for mineral resources is consequently likely to intensify even further the so-called second wave of the scramble for Africa, potentially intensifying conflicts in the region.  \\

\noindent  This new scramble for Africa is different from the first one in many aspects. One of them is that due to globalization, different actors and not only Europe have economic and geopolitical interests in the region. Chinese firms notably invest heavily on the continent \citep{carmody2017new}, but Western powers and other emergent countries are also involved \citep{ayers2013beyond}. Nonetheless, there are some important similarities, especially because resources  were crucial in the first scramble for Africa \citep{carmody2017new}.  \\

\noindent As argued above, natural resources have been shown to considerably matter for conflict and more generally violence. Therefore, conducting comprehensive measurements to determine the impact of mineral trade on conflict, as well as identifying which type of minerals is most pertinent to explain such a relationship, holds considerable importance from both short-term and long-term perspectives. This is particularly crucial in light of the imminent challenges posed by climate change.

\end{section}

\section{Data}

\subsection{Data construction}

I use trade data at the country level (UN Comtrade) between 2000 and 2019. This dataset contains information about exports/imports of all the types of products at different aggregation levels (HS2, HS4, HS6). The data is restricted to all the products that African countries exported to the rest of the world. A comprehensive description of the different types of minerals of interest is presented in Appendix \ref{app: commodities}. \\ 

\noindent Data from the CEPII is used to proxy prices using unit values \citep{berthou2011trade}. The authors argue that prices are not available for a large number of countries, products and years, which is why using trade unit values allows better empirical analysis of trade. The data used in this paper is the Trade Unit Values dataset and is constructed to "provide reliable and comparable unit values across countries" \citep{berthou2011trade}. \\ 

\noindent Conflict data comes from ACLED and reports information on the type, agents, location, date, and other characteristics of political violence events, demonstrations and select politically relevant non-violent events\footnote{ Armed Conflict Location \& Event Data Project (\url{https://acleddata.com/data-export-tool/})}. The relevant information for this study are the aggregated number of conflicts and fatalities. \\

\noindent Finally, both unemployment (used as a control) and GDP data (used to create the shares) come from the World bank open database\footnote{I used the STATA package \textit{wbopendata} which allows to download the data directly from STATA. More information can be found here: \url{https://github.com/jpazvd/wbopendata}}. 

\subsection{Data visualization}

\begin{figure}[h]
    \centering
    \caption{Conflicts and Mineral Trade in Africa \label{fig:fig1}}
    {{\includegraphics[width=17.5cm]{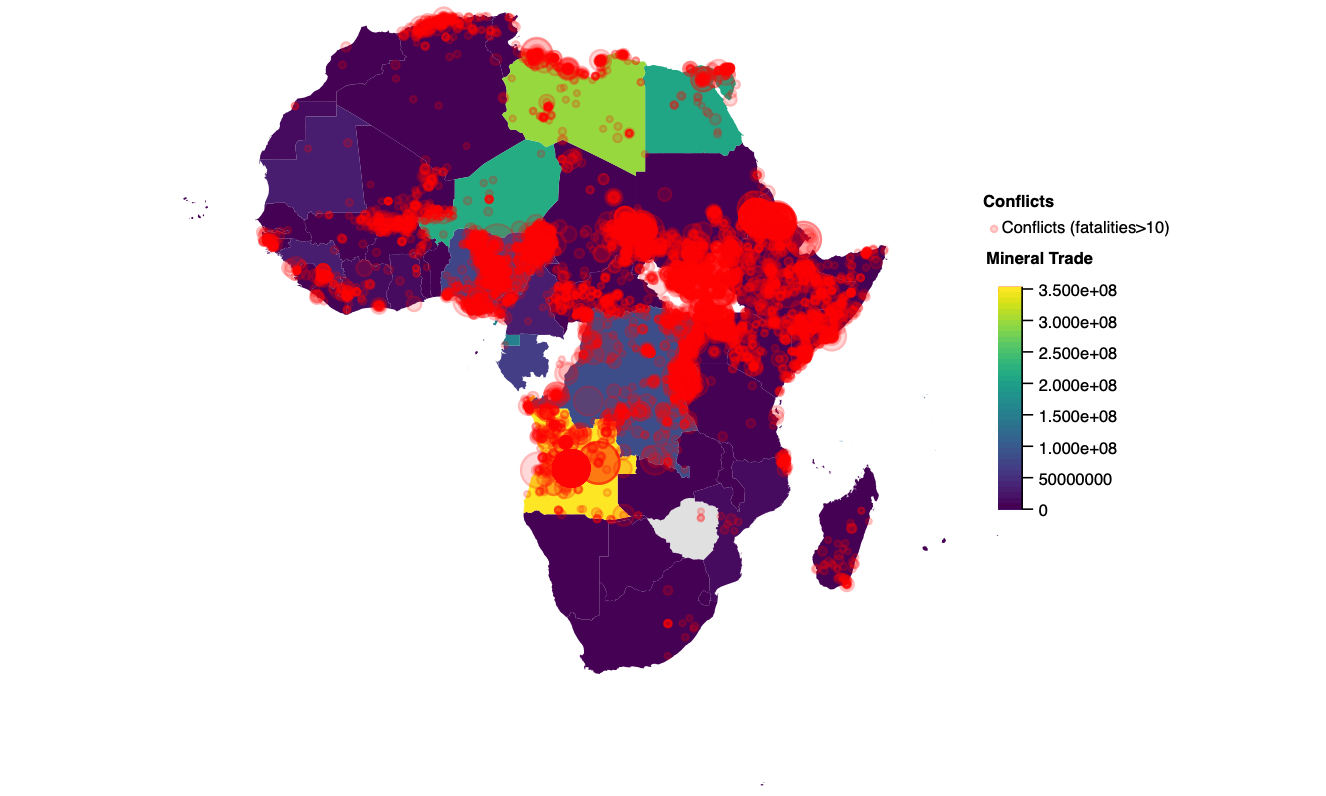} }}

    {\small{Note: This map presents the average mineral trade exports (in dollars) for each African country (between 2000 and 2020) and  the number of conflicts over the period (with more than 10 fatalities).}}

\end{figure}

This section provides a data visualization for the relationship between mineral trade and conflicts in Africa. Figure \ref{fig:fig1} displays the number of conflicts and the average mineral trade for each African country between 2000 and 2020. The data is restricted to conflicts resulting in more than 10 fatalities for readability purposes. Appendix figure \ref{fig: conflicts} presents the detailed map for all conflicts occurring in Africa over the period. \\

\noindent Notice that countries with high levels of mineral exports, such as Angola or the Democratic Republic of the Congo (DRC), experience a high frequency of fatal conflicts. While it remains impossible to establish direct causal relationships based solely on this correlation, it provides good motivating evidence to investigate the issue further. \\

\subsection{Descriptive statistics}

\noindent Table \ref{tab:summary stats} presents a summary statistics on three key outcome variables: incidence of a conflict, the number of conflicts and number of fatalities of the conflict. It also displays information on mineral trade values in US\$.  Observe that in the raw data, countries with a level of mineral exports higher than the median level have a higher incidence of conflicts on average. These countries also have a higher mean number of conflicts and fatalities. On the other hand, it is not as clear that countries exporting other commodities more than the median country have higher levels of conflict. Indeed, either the difference between below/above the median is not as large as in the case of minerals, or the relationship is even reversed in the case of fatalities. \\

\begin{table}[htbp]\centering
\def\sym#1{\ifmmode^{#1}\else\(^{#1}\)\fi}
\caption{Summary Statistics for Main Variables  \label{tab:summary stats}}
\begin{tabular}{l*{1}{cccccc}}
\toprule
                    &         &        Mean&          SD&         Min&         Max&           N\\
\midrule
Incidence \\ 

  \small{\textit{All Countries}}    &            &        0.89&        0.32&           0&           1&       1,252\\
   \small{\textit{Mineral Trade > Median}}    &            &        0.96&        0.20&           0&           1&         628\\
  \small{\textit{Mineral Trade < Median}}    &            &        0.81&        0.39&           0&           1&         624\\
   \small{\textit{Other commodities > Median}} &  &       0.91&        0.28&           0&           1&      11,596\\
   \small{\textit{Other commodities < Median}} &  &       0.86&        0.35&           0&           1&      11,596\\
    \midrule
  Number of conflicts \\
  \small{\textit{All Countries}}  &            &      184.41&      416.62&           1&       3,140&       1,109\\
 \small{\textit{Mineral Trade > Median}} &            &      223.13&      428.05&           1&       2,624&         602\\
  \small{\textit{Mineral Trade < Median}}&            &      138.44&      398.18&           1&       3,140&         507\\
  \small{\textit{Other commodities > Median}} &  &       212.27 &      415.84&           1&       3,140&      10,572\\
   \small{\textit{Other commodities < Median}} &  &       153.61&      384.84&           1&       3,140&       9,926\\
    \midrule
  Fatalities\\ 
  \small{\textit{All Countries}}  &            &      490.83&     1406.81&           0&      11,451&       1,109\\

  \small{\textit{Mineral Trade > Median}}    &            &      652.36&     1747.07&           0&      11,451&         602\\
 \small{\textit{Mineral Trade < Median}}    &            &      299.03&      800.33&           0&       6,138&         507\\
 \small{\textit{Other commodities > Median}} &  &       446.75&     1332.44&           0&      11,451&      10,572\\
   \small{\textit{Other commodities < Median}} &  &       491.84&     1370.53&           0&      11,451&       9,926\\
    \midrule
Mineral Trade \\
\small{\textit{All groups}} & & $3.57*10^7$ &    $1.32*10^8$ &       1004 &   $1.89*10^9$ &  2,553 \\
\small{\textit{Minerals 25}} & & 721519.5  &   1436001   &    1004  & $1.57*10^7$ & 934  \\
\small{\textit{Minerals 26}} & &    $1.47*10^7$ &   $3.54*10^7$   &    1143  & $3.47*10^8$ & 769 \\
\small{\textit{Minerals 27}} & &  $9.31*10^7$ &   $2.15*10^8$   &    1062 &  $1.89*10^9$ & 850 \\
\bottomrule
\end{tabular}

\vspace{0.2cm}
\textit{Note:} This table presents a summary statistics on conflict incidence (whether a conflict occurred in the country for a given year), the number of conflicts and fatalities (if a conflict occurred). It also provides information on mineral trade values in \$.  Note that the most aggregated level of analysis is used for this summary statistics, grouping every HS2 level related to minerals together.
\end{table}

\noindent In shift-share analysis, it is also interesting to include summary statistics for shocks, as they are the primary source of variation in the analysis. Table \ref{tab:Shock summary} reports summary statistics for shocks $g_{kt}$ computed with importance weights $s_{kt}$ (as in \cite{borusyak2022quasi}). Different specifications are presented, considering the entire sample and restricting the sample to minerals only. In both cases, the effective number of shocks, measured as the inverse of the Herfindahl index (HHI), ${1}/\sum_k s_{kt}$, is relatively high. On top on that, the largest shock weight is equal to 5.3\% using the full sample and 10\% using only minerals, suggesting a significant level of variation at the industry level. \\

\begin{table}[htbp]\centering
\def\sym#1{\ifmmode^{#1}\else\(^{#1}\)\fi}
\caption{Shock Summary Statistics}\label{tab:Shock summary}
\begin{tabular}{l*{4}{c}}
\hline
            &\multicolumn{1}{c}{(1)}&\multicolumn{1}{c}{(2)} &\multicolumn{1}{c}{(3)} &\multicolumn{1}{c}{(4)}\\
\hline
\textbf{Statistics} & & \\
Mean &  594646.4	 & 0 & 353.82 & 0   \\
Standard Deviation & 1831366 & 1741155 & 196.48 & 151.53 \\
Interquartile range & 4644.47 & 513825.1 & 247.61 & 195.94 \\
\textbf{Observation counts} & & \\
Number of industry-period shocks & 1479 & 1479 & 51 & 51  \\
Number of industries (HS2) & 87 & 87 & 3 & 3  \\
Effective number of shocks (1/HHI of $s_{kt}$ weights) &  71.75 & 71.75 & 23.10 & 23.10   \\
Largest $s_{kt}$ weight & 0.053 & 0.053 & 0.10 & 0.10 \\
\textbf{Specification} & & \\
Residualizing on period FE    &    No   &     Yes &  No & Yes   \\
All Industries & Yes & Yes & No & No \\
\hline
\end{tabular}
\\[0.3cm]
\small\textit{Note:} This table summarizes the distribution of price commodity shocks $g_{kt}$ across industries $k$ and periods $t$. All statistics are weighted by the average industry exposure shares $s_{kt}$. Shares are computed as the ratio of exports shares in the industry to GDP and are lagged by 3 periods. Columns (1) and (2) include all industries while columns (3) and (4) restrict the sample to mineral industries only (using the HS2 level). Column (2) and (4) also residualizes the shocks on period fixed effects. I also report the effective number of shocks which is the inverse of the Herfindahl index of shocks importance weights, ${1}/\sum_k s_{kt}$.
\end{table}

\noindent An additional summary statistics on which countries are the most important for mineral trade and to what extent countries in the top 10 of mineral exporters contribute to trade of mineral products in presented in Appendix Table \ref{app:percent}. This will be important when discussing the identification assumptions for the shift-share analysis based on exogenous shocks.

\hspace{4cm}

\section{Methodology}

 \subsection{Country level specification}

I want to estimate the following equation: 

\begin{equation}
    y_{lt} = \beta_0 + \beta_1 x_{lt} + \beta_2 w_{lt} + \delta H_{lt} + \psi_{t} + \eta_{l} + \varepsilon_{lt}
\end{equation}

\noindent $y_{lt}$ represents the conflict variable (number of conflicts, fatalities etc) for country $l$ at time $t$. $x_{lt}$ is the treatment variable: exports of minerals (in base period USD). The idea behind the treatment is to interpret exports of minerals as a demand shock to investigate the hypothesis that a higher demand for minerals might have a social cost in Africa because of an increase in conflicts. $w_{lt}$ controls for the exports of all other products and $H_{lt}$ controls for unemployment. $\psi_{t}$ and $\eta_{l}$ represent time and country fixed-effects, respectively. \\

\noindent As $x_{lt}$ is likely endogenous, I use a SSIV methodology.
A set of exogenous shocks $g_{kt}$ for $k \in \{1,...K\}$ mineral industries and shares $s_{lkt} \geq 0$ are used to estimate $\beta_1$.
The shocks are worldwide prices of the products  while shares are defined as export shares as a fraction of GDP: 

\begin{equation}
s_{lkt} = \frac{X_{lkt}}{GDP_{lt}}
\end{equation} 

\noindent Doing so would potentially give the instrument more power (a stronger first stage) than just using exports of country $l$ in industry $k$ and year $t$ divided by total exports of country $l$ in year $t$. Indeed, the instrument should have more variation in countries where the export to GDP ratio is large, which is not captured by just using total exports. Note that since the shares do not add up to one in that case, I control for the share of total exports in GDP, interacted with period effects (referred as the "incomplete share control" in \cite{borusyak2022quasi}). The control is defined as: 

\begin{equation}
S_{lt} = \frac{X_{lt}}{GDP_{lt}}
\end{equation}

\noindent The instrument is hence the exposure-weighted average of the shocks:

\begin{equation}
    z_{lt} = \sum_{k=1}^K s_{lkt} g_{kt}
\end{equation} 

\noindent Note that this setup follows the methodology proposed by  \cite{borusyak2022quasi}, where shocks are exogenous while shares can be endogenous. The first and second stage equations can be expressed as follows: 

\begin{equation}
    \hat{x}_{lt} = \gamma_0 + \gamma_1 z_{lt} + \gamma_2 w_{lt} + \delta H_{lt} + \psi_{t} + \eta_{l} + \lambda S_{lt} * \psi_{t} +\varepsilon_{lt}
\end{equation}

\begin{equation} \label{2nd stage}
y_{lt} = \beta_0 + \beta_1 \hat{x}_{lt} + \beta_2 w_{lt} + \delta H_{lt} + \psi_{t} + \eta_{l} + \lambda S_{lt} * \psi_{t} + \varepsilon_{lt}
\end{equation}

\noindent To avoid the bad controls problem,  $w_{lt}$ is instrumented analogously but summing through all industries but minerals, which yields  a just-identified IV regression with two endogenous variables. I discuss the implications of having multiple endogenous variables in Appendix \ref{appendix: 2 endog}.

\subsection{Industry level specification} 

\noindent \cite{borusyak2022quasi} show that $\hat{\beta}$ is
equivalently obtained as the coefficient from a non-standard shock-level IV procedure, in which $g_k$ directly serves as the instrument. Indeed, the SSIV estimator of $\beta_1$ in Equation \ref{2nd stage} equals the second stage estimator of a shares-weighted IV regression that uses the shocks $g_n$, in this case prices, as an the instrument in estimating 

\begin{equation}
\bar{y}_{kt}^\perp = \beta_0 + \beta_1 \bar{x}_{kt}^\perp + \bar{\varepsilon}_{kt}^\perp
\end{equation} 

\noindent where $v_{lt}^{\perp}$ denotes the in-sample projection of $v_{lt}$ on $H_{lt}$ for any variable $v_{lt}$ on the control vector $X_{lt}$ and $\bar{v}_{kt} = \frac{\sum_l s_{lk}v_l}{\sum_l s_{lk}}$, following notation from \cite{borusyak2022quasi}. \\

\noindent Running the specification at the industry level allows to compute exposure robust standard errors. Indeed, since the shock is at the industry level, countries cannot be viewed as \textit{iid} observations.  \cite{adao2019shift} argue that inference must take into account that units with relatively similar shares mechanically have correlated instruments and may also be exposed to the same unobserved shocks and have correlated $\varepsilon_i$. In this case, countries specializing in the same industries will be affected by the same price shocks. An advantage of using shock-level regression is that it allows to cluster by groups of industries for instance, yielding valid standard errors \citep{borusyak2022quasi}.

\subsection{Identification}\label{Identification}

The standard shift-share IV identification assumptions apply to this context. First, the instrument should be relevant (have power) i.e. $\mathbb{E}[x_{lt} z_{lt} | w_{lt}, H_{lt}] \neq 0$. This condition can be checked in the data by computing the first-stage F statistic, but it is most likely that price of minerals affects trade of minerals.  For the exclusion restriction, \citep{borusyak2022quasi}  argue that the following conditions are equivalent

\begin{equation}
\mathbb{E} \left[ \sum_l{z_l \varepsilon_l}\right]
 =0 \iff \mathbb{E}\left[\sum_n{s_n g_n \bar{\varepsilon}_l}\right]=0,
\end{equation}

\noindent where $s_k = \frac{1}{L }\sum_l s_{lk}$ and $\bar{\varepsilon}_l = \frac{\sum_l s_{lk}\varepsilon_l}{\sum_l s_{lk}}$. The first expression captures the orthogonality of the instrument $z_l$ with the residual $\varepsilon_l$ while the second equality captures the orthogonality of the shock $g_k$ with $\bar{\varepsilon}_l$, weighted by $s_k$. Hence, by equivalence, the latter is a necessary and sufficient condition for the orthogonality of the shift-share instrument \citep{borusyak2022quasi}. In this context, the shocks (worldwide prices of minerals) need to be exogenous meaning that countries of interest are not able to determine prices of commodities. \\

\noindent A potential threat to identification would be if some big mineral producers\footnote{An important African mineral exporter that directly comes to mind is the DRC as they are the main producers of minerals such as cobalt.} were able to produce enough to influence the world market price, leading to endogeneity and potential failure of the exclusion restriction. The structure of the data nonetheless allows to alleviate these concerns. Indeed, since minerals\footnote{A selection of minerals is presented in Appendix Table \ref{app:Description}} are still measured at a pretty high level of aggregation, the mineral exports variables contain different types of minerals, making it less likely for their price to be solely determined by one country. \\

\noindent Appendix Table \ref{app:percent} presents an additional descriptive statistics providing evidence that mineral prices are likely to be exogenous in this context. This table presents the top 10 exporters of minerals for years 2000 and 2020, reporting how much they contribute in percent to total mineral exports during this year. Even in the top 10 of exporters, countries contribute generally to less than 10\% of mineral exports in a year and no African country is in the top 10 of mineral exporters in the world. \\

\noindent This provides a strong case for the exclusion restriction as it highlights that prices of minerals can be considered as exogenous to export decisions of African countries, meaning that the SSIV design based on exogenous shocks seems to be a valid approach. \\

\newpage

\section{Results}\label{sec5}

\subsection{Main results}

This section presents the main results of this study on the causal effect of mineral trade on two outcome variables: number of conflicts and number of fatalities. Three different specifications are presented. First, a basic OLS regression with fixed effects and clustering at the industry level. Second, a SSIV specification with fixed effects but without clustering. Finally, the preferred specification is the SSIV clustering at the industry level. Since shocks are serially correlated, shares are lagged for three periods and I control for the lagged outcome variable \citep{borusyak2022quasi}. \\

\noindent First of all, notice that the F-statistics for the first stage is higher than 10, a confirmation that the relevance condition is satisfied and that the problem of weak instrument can be ruled out. The most conservative estimator shows that an increase of 1 million of dollars  of mineral trade increases conflicts by $0.059$. The coefficient is statistically significant at the $10\%$ level. Note that  clustering at the industry level only affects the standard errors but not the value of the coefficient. \\

\noindent To understand the magnitude of this coefficient better, note that one standard deviation in the mineral trade values in an African country for one year between 2000 and 2020 equals approximately 132 million dollars (see Table \ref{tab:summary stats}). An increase of one standard deviation in mineral trade hence corresponds to an increase of almost 8 conflicts, which is not negligible.   \\

\noindent When it comes to the number of fatalities, results are less readable. Indeed, it seems that clustering at the industry level makes the coefficient insignificant. On top of that, when performing the falsification test in subsection \ref{robustness}, the coefficient for lagged fatalities is significant and different from zero. It is therefore hard to interpret this coefficient as causal. \\

\noindent Finally, observe that the coefficients on non-mineral trade are negative and significant for number of conflicts, potentially pointing to the fact that being well integrated in the globalized economy leads to less conflicts. This would be consistent with the liberal view of international relations advocating that globalization and the spread of free market should reduce the use of military force among states \citep{thoenig2023trade}. 

\begin{table}[H]\centering
\def\sym#1{\ifmmode^{#1}\else\(^{#1}\)\fi}
\caption{Causal Effect of Mineral Trade on Conflicts with SSIV}\label{tab: Main Results}
\begin{tabular}{l*{3}{c}}
\hline
            &\multicolumn{1}{c}{(1)}&\multicolumn{1}{c}{(2)}&\multicolumn{1}{c}{(3)}\\
            &\multicolumn{1}{c}{Conflicts 1}&\multicolumn{1}{c}{Conflicts 2}&\multicolumn{1}{c}{Conflicts 3}\\
\hline
Mineral Trade &        0.007** &       0.059***&       0.059*    \\
             &    (0.003)   &     (0.012)   &     (0.033)    \\
[1em]
Non Mineral Trade  &    0.053*** &      -0.356***&      -0.356*** 
\\
& (0.006)   &     (0.076)   &     (0.107) \\
[1em]
F-test    &         &      60.976   &      10.510      \\
Controls    &      Yes   &      Yes   &      Yes     \\
FE    &      Yes   &      Yes   &      Yes   \\
Clustering & Yes & No & Yes \\
\hline\hline
            &\multicolumn{1}{c}{(1)}&\multicolumn{1}{c}{(2)}&\multicolumn{1}{c}{(3)}\\
            &\multicolumn{1}{c}{Fatalities 1}&\multicolumn{1}{c}{Fatalities 2 }&\multicolumn{1}{c}{Fatalities 3}\\
\hline
Mineral Trade &       0.015   &      -0.078*** &      -0.059   \\
            &     (0.017)   &     (0.015)   &     (0.048)         \\
[1em]
 Non Mineral Trade  &   0.053***&      -0.044   &      -0.064   \\

& (0.019)   &     (0.098)   &     (0.132)   \\
[1em]

F-test     &         &      60.976   &      52.746    \\
Controls    &      Yes   &      Yes   &      Yes     \\
FE    &      Yes   &      Yes   &      Yes     \\
Clustering & Yes & No & Yes \\
\hline
\multicolumn{4}{l}{\small \sym{*} \(p<0.1\), \sym{**} \(p<0.05\), \sym{***} \(p<0.01\)}\\
\hline
\end{tabular}
\\[0.3cm]
\small\textit{Note:} This table presents the SSIV estimates showing the impact of trade of minerals on conflicts. \\
Column (1) presents the OLS estimator, Column (2) the SSIV estimator without clustering and Column (3) the SSIV estimator clustering at the industry level. The F-test corresponds to the first stage F-test.  Both SSIV specifications control for lagged outcome. Note that exports are normalized by 1'000'000 for readability of the results. 
\end{table}

\subsection{Robustness} \label{robustness}

Table \ref{tab:Fals} presents a falsification test using lagged outcome variable, as in \cite{aghion2020labor}. In the first two columns of Table \ref{tab:Fals}, I cannot reject
that there is no relationship between mineral trade (instrumented with the shift-share instrument) and lagged number of conflicts.  The point estimates are also smaller in magnitude than those from Table \ref{tab: Main Results}. Overall, this falsification test is reassuring for the robustness of the results for number of conflicts. \\

\noindent On the other hand, the number of fatalities does not "pass" the falsification test when fatalities are lagged at period $t-1$. Indeed, the coefficient is highly significant and the diverges from zero. Coupling that with the fact the results for fatalities were not robust to clustering, it is hard to claim any causal effect of mineral trade on the number of fatalities, at least at such a high level of aggregation. \\

\begin{table}[htbp]\centering
\def\sym#1{\ifmmode^{#1}\else\(^{#1}\)\fi}
\caption{Falsification Test}\label{tab:Fals}
\begin{tabular}{l*{4}{c}}
\hline
            &\multicolumn{1}{c}{(1)}&\multicolumn{1}{c}{(2)}&\multicolumn{1}{c}{(3)}&\multicolumn{1}{c}{(4)}\\
            &\multicolumn{1}{c}{Conflicts (t-1)}&\multicolumn{1}{c}{Conflicts (t-2)}&\multicolumn{1}{c}{Fatalities (t-1)}&\multicolumn{1}{c}{Fatalities (t-2)}\\
\hline
Mineral Trade&       0.009   &       0.010   &      -0.111***&       0.051   \\
            &     (0.015)   &     (0.016)   &     (0.043)   &     (0.132)   \\
[1em]
F-test     &   30858   &    9198   &   30858   &    9198   \\
Controls    &      Yes   &      Yes   &      Yes   &      Yes  \\
FE    &      Yes   &      Yes   &      Yes   &      Yes  \\
Clustering  &   Yes   &      Yes   &      Yes   &      Yes  \\
\hline
\multicolumn{5}{l}{\small \sym{*} \(p<0.1\), \sym{**} \(p<0.05\), \sym{***} \(p<0.01\)}\\
\hline
\end{tabular}
\\[0.3cm]
\small\textit{Note:} This falsification test table presents the SSIV estimates showing the impact of trade on lagged conflict outcomes variables (number of conflict and fatalities) at periods t-1 and t-2.  The F-test corresponds to the first stage F-test.
\end{table}

\section{Heterogeneity}\label{sec:heterog}

The goal of this section is to investigate further which type of mineral is important when explaining conflict.  Distinguishing between the three types of mineral products at the HS2 level allows to verify if the results from section \ref{sec5} are driven by some specific groups of minerals. I estimate the following just-identified SSIV regression

\begin{equation}\label{eq: heterog}
y_{lt} = \beta_0 + \beta_1 x_{1lt} + ... + \beta_R  x_{Rlt} + \beta_{R+1} w_{lt} + \delta H_{lt} + \varepsilon_{lt}
\end{equation}

\noindent where $x_{1lt},...,x_{Rlt}, w_{lt}$ are instrumented by a set of SSIV and $z_{1lt},...,z_{R+1lt}$ where $z_{rlt} = \sum_{k=1}^K s_{rlkt}g_{rkt}$. $x_{rlt}$ represents the trade of mineral $r$ for country $l$ at time $t$ while $w_{lt}$ controls for exports of other commodities as before. $H_{lt}$ collects all other controls (and fixed effects). 
This corresponds to an SSIV regression with $R+1$ endogenous variables. More details can be found in Appendix \ref{appendix: R+1 endog}. \\

\noindent Table \ref{tab: heterogeneity} provides the results for the heterogeneity analysis at the HS2 level. The HS nomenclature creates three subgroups in the mineral industry: categories 25, 26, 27.  More information about the groups and examples can be found in Appendix \ref{app: commodities}. I estimate Equation \ref{eq: heterog} using these three subgroups as endogenous variables, instrumented by their respective shift-share instruments.  Two specifications (clustering at the industry level or not) are performed for the same two outcome variables as before (number of conflicts and number of fatalities). I only interpret significant and robust results here (i.e. results that are significant in both specifications). \\

\noindent First, it is interesting to highlight that the only type of conflict that seems to matter to explain the number of conflicts is the category 27, which contains mainly mineral fuels/oils such as gas or petrol. Indeed, an increase of 1 million dollars in mineral type 27 exports increases conflicts by 0.064. The coefficient is the same with or without clustering at the industry level, and remains significant at the 10\% level while clustering.   \\

\noindent Note that the standard deviation of trade of mineral type 27 is approximately 215 million dollars, implying that an increase of one standard deviation in trade of this specific mineral type yields an increase of almost 14 conflicts.  Notably, it appears that despite the observed rise in conflicts associated with this particular type of mineral, there is a
decrease in fatalities. This suggests that conflicts become less lethal when trade in this mineral group intensifies. \\

\noindent Moreover, and potentially even more interesting, it seems that an increase in trade of mineral type 26 (which contains minerals such as coltan, precious metals, cobalt etc.) does not increase the number of conflicts but has a high statistically significant coefficient of 0.362 on fatalities. The standard deviation of this mineral type is around 35 million dollars, which implies that an increase of one standard deviation in the trade of mineral type 26 increases number of fatalities by almost 13.  \\

\noindent Overall, these results are consistent with the hypothesis that an increase in demand of rare natural resources leads to highly fatal conflicts and a general increase in violence. Note that the impact of mineral trade on fatalities was not clear in the initial - most aggregated - analysis, while it appears that a specific subgroup of (mainly precious) minerals seems to extensively matter, highlighting that conducting such heterogeneity analysis in this context can be extremely insightful.  \\

\noindent One potential avenue for deeper analysis could be to incorporate each individual mineral at the HS4 level into Equation \ref{eq: heterog}. However, due to the limited explanatory power, such an approach would likely yield uninteresting results. This study can thus be seen as an endeavor to investigate which sub-group of mineral matters, narrowing the focus for further research and policy considerations.

\begin{table}[H]\centering
\def\sym#1{\ifmmode^{#1}\else\(^{#1}\)\fi}
\caption{Heterogeneity Results}\label{tab: heterogeneity}
\begin{tabular}{l*{4}{c}}
\hline
            &\multicolumn{1}{c}{(1)}&\multicolumn{1}{c}{(2)}&\multicolumn{1}{c}{(3)}&\multicolumn{1}{c}{(4)}\\
            &\multicolumn{1}{c}{Conflicts 1}&\multicolumn{1}{c}{Conflicts 2}&\multicolumn{1}{c}{Fatalities 2}&\multicolumn{1}{c}{Fatalities 2}\\
\hline
Mineral Trade (25)&      -0.461   &      -0.461   &       2.360***&       2.360   \\
            &     (0.418)   &     (0.913)   &     (0.719)   &     (1.534)   \\
[1em]
Mineral Trade (26) &      -0.098   &      -0.098   &       0.362***&       0.362*  \\
            &     (0.069)   &     (0.125)   &     (0.119)   &     (0.201)   \\
[1em]
Mineral Trade (27)&       0.064***&       0.064*  &      -0.110***&      -0.110*  \\
            &     (0.012)   &     (0.034)   &     (0.020)   &     (0.056)   \\
[1em]
Non Mineral Trade &      -0.224***&      -0.224** &       0.042   &       0.042   \\
            &     (0.052)   &     (0.106)   &     (0.089)   &     (0.142)   \\
[1em]

F-test     &   186.598   &   11.220 &     186.598   & 11.220            \\
Controls    &      Yes   &      Yes   &      Yes   &      Yes  \\
FE    &      Yes   &      Yes   &      Yes   &      Yes  \\
Clustering  &   No   &      Yes   &      No   &      Yes  \\
\hline
\multicolumn{5}{l}{\small \sym{*} \(p<0.1\), \sym{**} \(p<0.05\), \sym{***} \(p<0.01\)}\\
\hline
\end{tabular}
\\[0.3cm]
\small\textit{Note:}  This table presents the SSIV estimates showing the impact of trade of minerals on conflicts, separated in 3 distinct types of minerals. Column (1) presents the SSIV estimator for the number of conflicts without clustering and column (2) when clustering at the industry level.  Columns (3) and (4) reproduce the same for the number of fatalities. The F-test corresponds to the first stage F-test.  Both SSIV specifications control for lagged outcome. Exports are normalized by 1'000'000 for readability of the results. 
\end{table}

\section{Discussion and Limitations}

This section presents some potential limitations of this paper. First, as briefly mentioned at the end of part \ref{sec:heterog}, the sample size does not allow to conduct the analysis at the HS4 level, as this would imply running equation \ref{eq: heterog} with approximately 60 endogenous variables, which would considerably affect the explanatory power of the model. Some further research could be done to create an \textit{artificial HS3 category}, grouping minerals in a sensible and less aggregated manner than at the HS2 level, but without having 60 different categories. Such groups might be perceived as arbitrary, but could still provide some information to obtain a better and more structured understanding of which specific minerals can heavily impact conflicts in Africa. \\

\noindent On top of that, even if we discussed in both sub-section \ref{Identification} and Appendix \ref{app:Description} that African countries don't seem to affect worldwide prices too much, they are still part of the system which implies that prices are not completely exogenous. This does not mean that the exclusion restriction fails, as it seems that African countries are not able to control the market price, but it would be interesting to use another source of variation which is completely exogenous to Africa (for instance shocks happening in other continent that plausibly have nothing to do with the African mineral market) and see whether the results are robust to this alternative method. This would be even more important when going to a further level of disaggregation where some specific countries might be able to impact commodity prices heavily, in the potential \textit{HS3 category} for instance. \\

\noindent Finally, even if shares can be endogenous in a SSIV design based on exogenous shocks, it could be interesting to use exposure shares that are less endogenous than the ratio of mineral trade to GDP. For instance, data on mineral reserves (a segment of a mineral resource that has undergone comprehensive evaluation and is considered commercially viable\footnote{\url{https://www2.bgs.ac.uk/mineralsuk/mineralsYou/resourcesReserves.html}}) could potentially provide less endogenous shares as they would be determined (at least partly) naturally.

\section{Conclusion}

This paper uses a SSIV methodology based on an exogenous price-commodity shock weighted by mineral industry shares to measure the  impact of mineral trade on conflict in
Africa between 2000 and 2020. The first result of this study is that an increase of one standard deviation in mineral trade corresponds to an increase of almost 8 conflicts. This finding is robust to different specifications and falsification tests. On the other hand, it seems that mineral trade (as a whole) has no statistically significant effect on fatalities. \\

\noindent On top of that, the heterogeneity analysis shows that different groups of minerals affect conflict outcomes in a differentiated manner. First, category 27, which contains mainly mineral fuels/oils such as gas or petrol is the only group increasing the number of conflicts. An increase of one standard deviation in trade of this specific mineral type yields an increase of almost 14 conflicts. Moreover, an increase of one standard deviation in the trade of mineral type 26 increases number of fatalities by almost 13, while it appears to have no effect on the number of conflicts. \\

\noindent The finding that an increase in trade of minerals such as cobalt, precious metals or coltan increases  fatalities could provide an explanation of why some African regions, notably the East part of the continent, are experiencing high degrees of violence and deaths due to conflict.  \\

\noindent By adopting a novel estimation strategy in this context (SSIV), this study contributes to the existing body of literature on natural resources trade and conflicts. While the heterogeneity results remain relatively aggregated, they pave the way for future research endeavors aimed at achieving a thorough and comprehensive understanding of the specific minerals that significantly influence conflicts in Africa. Such insights hold direct policy implications, particularly in the context of addressing the challenges posed by climate change.

\newpage

\printbibliography

\appendix
\setcounter{table}{0}
\renewcommand{\tablename}{Appendix Table}
\renewcommand{\figurename}{Appendix Figure}
\renewcommand{\thetable}{A\arabic{table}}
\setcounter{figure}{0}
\renewcommand{\thefigure}{A\arabic{figure}}
\newpage

\section{Appendix}
\subsection{Commodities description and exporter importance} \label{app: commodities}

This section provides information on the different groups of minerals used throughout the paper. Following the HS nomenclature, 3 categories of commodities are categorized as mineral products: HS25, HS26, HS27\footnote{Most information can be found here: \url{https://www.wcoomd.org/en/topics/nomenclature/instrument-and-tools/hs-nomenclature-2022-edition/hs-nomenclature-2022-edition.aspx}}. Category HS25, called \textbf{salt, sulphur, earths and stone, plastering materials, lime and cement}, contains only products which are in the crude state or which have been washed, but not products which have been roasted, calcined, obtained by mixing or subject to processing. Category HS26 is called \textbf{ores, slag and ash} and contains mostly what one think about when hearing minerals. Finally, category 27 is referred as \textbf{mineral fuels, mineral oils and products of their distillation, bituminous substances, mineral waxes}. Examples are provided in Appendix Table \ref{app:Description}.

\begin{center}
\begin{table}[htbp]\centering
\centering
\caption{Description of a select type of industries \label{app:Description}}
\begin{tabular}{lccc}
\cmidrule{1-4}
\begin{tabular}[c]{@{}l@{}}\end{tabular} &\begin{tabular}[c]{@{}l@{}}HS2\end{tabular} & \begin{tabular}[c]{@{}c@{}} HS4  \end{tabular} & \begin{tabular}[c]{@{}c@{}} Description \end{tabular}  
\\                      \\
\cline{2-4} & (1) & (2) & (3)
                                            \\
\cmidrule{1-4}  
& 25	& 2502	& Unroasted iron pyrites  \\
& 25	& 2504	& Natural graphite  \\
& 25	& 2506	& Quartz  \\
& 25	& 2510	& Natural calcium phosphates  \\
& 26	& 2601	& Iron  \\
& 26	& 2602	& Manganese  \\
& 26	& 2603	& Copper  \\
& 26	& 2604	& Nickel  \\
& 26	& 2605	& Cobalt  \\
& 26	& 2606	& Aluminium   \\
& 26	& 2607	& Lead   \\
& 26	& 2608	& Zinc   \\
& 26	& 2609	& Tin   \\
& 26	& 2615	& Niobium, tantalum, vanadium or zirconium   \\
& 27	& 2701	& Coal \\
& 27	& 2707	& Oils and other products of the distillation of high temperature coal tar \\
& 27	& 2709	& Petroleum oils and oils obtained from bituminous minerals \\
& 27	& 2711	& Petroleum gases and other gaseous hydrocarbons \\
\hline
\end{tabular}
\label{tab6}
\\[0.3cm]
\small\textit{Note:} This table presents the description of different types of minerals codes.
\end{table} 
\end{center}

\noindent Appendix Table \ref{app:percent} provides an additional descriptive statistics showing why it is likely that the exclusion restriction (i.e. that world prices of mineral are exogenously determined and not affected by specific African countries) holds in this context. \\

\noindent This table presents the top 10 exporters of minerals for years 2000 and 2020, reporting how much they contribute in percent to total mineral exports during this year. Observe that except from Saudi Arabia in 2000, no country contributes  more than 10\% to the total mineral exports, and most countries of the top 10 contribute less than 5\%.  Moreover, it is interesting to note that no African country is in the top 10 of mineral exporters in the world. As mentioned above, it provides a strong case for the exclusion restriction as it highlights that prices of minerals can be considered as fully exogenous to export decisions of African countries, meaning that the SSIV design based on exogenous shocks seems to be a valid approach. 

\noindent 

\begin{center}
\begin{table}[htbp]\centering
\centering
\caption{Percentage of mineral export of the top 10 mineral exporters in 2000 and 2020\label{app:percent}}
\begin{tabular}{lccc}
\cmidrule{1-4}
\begin{tabular}[c]{@{}l@{}}\end{tabular} &\begin{tabular}[c]{@{}l@{}}Year\end{tabular} & \begin{tabular}[c]{@{}c@{}} Country  \end{tabular} & \begin{tabular}[c]{@{}c@{}} Share of Mineral Exports \end{tabular}

                                            \\
\cmidrule{1-4}

& 2000 & IRN & 2.24 \\
& 2000 & NLD & 3.07 \\
& 2000 & NGA & 3.32 \\
& 2000 & GBR & 3.75 \\
& 2000 & VEN & 3.86 \\
& 2000 & ARE & 4.01 \\
& 2000 & NOR & 5.20 \\
& 2000 & CAN & 5.52 \\
& 2000 & RUS & 6.84 \\
& 2000 & SAU & 10.20 \\
\hline

& 2020 & QAT & 2.54 \\
& 2020 & MYS & 2.55 \\
& 2020 & IRQ & 3.23\\
& 2020 & BRA & 3.70\\
& 2020 & CAN & 4.93 \\
& 2020 & ARE & 4.52 \\
& 2020 & SAU & 7.09 \\
& 2020 & USA & 8.13 \\
& 2020 & RUS & 8.90 \\
& 2020 & AUS & 9.83 \\
\hline
\end{tabular}
\\[0.3cm]
\small\textit{Note:} This table presents the percentage of mineral exports (calculated as the ratio of the countries' exports to total exports in the year) for the top 10 exporters in years 2000 and 2020. 
\end{table} 
\end{center}

\subsection{Additional visualization}
\begin{figure}[h]
    \centering
    \caption{Conflicts in Africa \label{fig: conflicts}}
    {{\includegraphics[width=14cm]{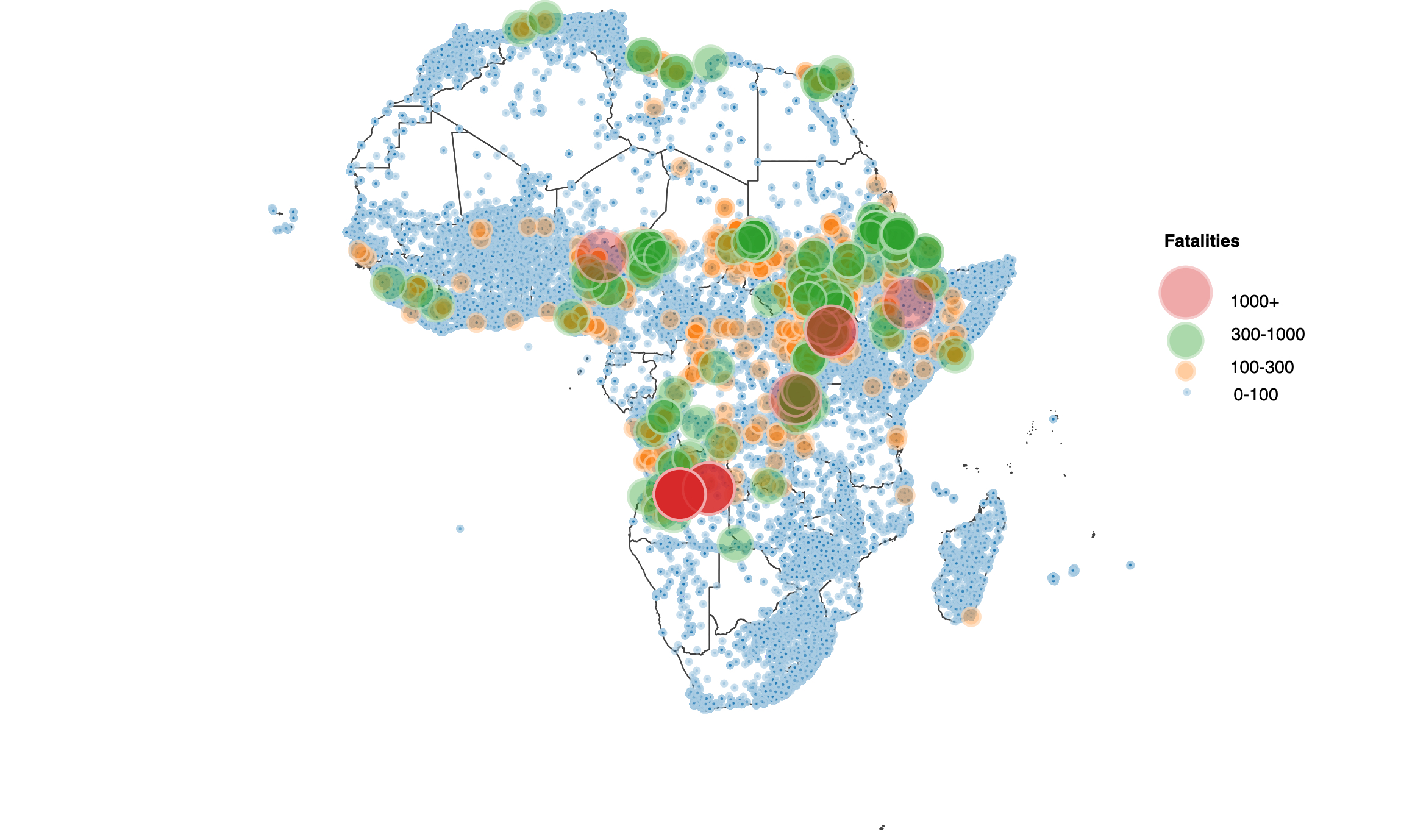} }}

    \small{Note: This maps presents all the conflicts happening in Africa between 2000 and 2020.}
\end{figure}

\subsection{SSIV Regression with two endogenous variables}\label{appendix: 2 endog}

Consider the just-identified SSIV regression 

\begin{equation}\label{2 endog}
    y_{lt} = \beta_0 + \beta_1 x_{lt} + \beta_2 w_{lt} + \delta H_{lt} + \varepsilon_{lt}
\end{equation}

\noindent where $x_{lt}$ is instrumented by $z_{1lt}$ and $w_{lt}$ is instrumented by $z_{2lt}$ where $z_{rlt} = \sum_{k=1}^K s_{rlkt}g_{rkt}$, $ \forall r \in\{1,2\}$.  $H_{lt}$ collects all other controls (and fixed effects), $s_{rlkt}$ represents the shares of the exports of industry $k$ in country $l$ at time $t$ for endogenous variable $r$ while $g_{rkt}$ is the shock variable.  Note that both $x_{lt}$ and $w_{lt}$ have the same number of shocks K since both shocks $g_{1kt}$ and $g_{2kt}$ are prices in the industry at time $t$. Following the Practical guide to SSIV\footnote{Still a working paper from Borusyak Jaravel and Hull (no citation available yet).}, let $v_{lt}^{\perp}$ denote the in-sample projection of $v_{lt}$ on $H_{lt}$ for any variable $v_{lt}$ ($y_{lt}, x_{lt}, w_{lt}$ in this case). 

\noindent Moreover, let $\tilde{v}_{kt}^{(r)} = \frac{1}{NT} \sum_{l,t} s_{rlkt} v_{lt}.$ \\

\noindent The IV estimator $\hat{\beta}$ for $\beta = (\beta_1, \beta_2)'$ in Equation \ref{2 endog} satisfies a system of $R=2$ equations 

\begin{equation}
   \frac{1}{NT} \sum_{l,t}\left(y_{lt}^{\perp} - \hat{\beta}_1x_{lt}^{\perp} - \hat{\beta}_2w_{lt}^{\perp} \right)z_{rlt} = 0
\end{equation}

\noindent The authors show that this expression is equivalent to 

\begin{equation}
      \sum_k \left(\tilde{y}_{kt}^{(r)} - \hat{\beta}_1\tilde{x}_{kt}^{(r)} - \hat{\beta}_2\tilde{w}_{kt}^{(r)} \right)g_{rkt} = 0
\end{equation}

\noindent which is nothing but two moments conditions saying that the shocks (prices) should be orthogonal to the residuals for both endogenous variables. 

\subsection{SSIV Regression with many endogenous variables}\label{appendix: R+1 endog}

As in Appendix \ref{appendix: 2 endog}, the IV estimator $\hat{\beta}$ for $\beta = (\beta_1,... \beta_{R+1})'$ in Equation \ref{eq: heterog} satisfies a system of $R+1$ equations

\begin{equation}
   \frac{1}{NT} \sum_{l,t}\left(y_{lt}^{\perp} - \sum_{j=1}^{R+1}\hat{\beta}_jx_{jlt}^{\perp}  \right)z_{rlt} = 0
\end{equation}

\noindent which is equivalent to the moment conditions 

\begin{equation}
      \sum_k \left(\tilde{y}_{kt}^{(r)} - \sum_{j=1}^{R+1}\hat{\beta}_j\tilde{x}_{jkt}^{(r)} \right)g_{rkt} = 0
\end{equation}

\newpage

\end{document}